\documentclass[lettersize,journal]{IEEEtran}
\usepackage{amsmath,amsfonts}
\usepackage{stfloats}
\usepackage{graphicx}
\usepackage{cite}
\usepackage{hhline}
\usepackage{booktabs}

\IEEEtriggeratref{27}

\begin{document}
\bstctlcite{BSTcontrol} 

\title{Vulnerability to Parameter Spread in\\ Josephson Traveling-Wave Parametric Amplifiers}

\author{C. Kissling, V. Gaydamachenko, F. Kaap, M. Khabipov, R. Dolata, A. B. Zorin, L. Grünhaupt
\thanks{Manuscript received November 18, 2022; accepted January 18, 2023.}
\thanks{C.K. and F.K. gratefully acknowledge the support of the Braunschweig International Graduate School of Metrology B-IGSM and the DFG Research Training Group 1952 Metrology for Complex Nanosystems. 
This work was also supported by the German Federal Ministry of Education and Research (BMBF)
within the framework programme “Quantum technologies -- from basic research to market” (Grant No. 13N15949).}%
\thanks{The authors are with Physikalisch-Technische Bundesanstalt, Bundesallee 100, 38116 Braunschweig, Germany (e-mail: christoph.kissling@ptb.de).}}%



\maketitle

\begin{abstract}
We analyze the effect of circuit parameter variation on the performance of Josephson traveling-wave parametric amplifiers (JTWPAs). 
Specifically, the JTWPA concept we investigate is using flux-biased nonhysteretic rf-SQUIDs in a transmission line configuration, which harnesses the three-wave mixing (3WM) regime. Dispersion engineering enables phase-matching to achieve power gain of $\sim$20~dB,  while suppressing the generation of unwanted mixing processes. 
Two dispersion engineering concepts using a 3WM-JTWPA circuit model, i.e., resonant phase-matching (RPM) and periodic capacitance modulation (PCM), are discussed, with results potentially also applicable to four-wave-mixing (4WM) JTWPAs. 
We propose suitable circuit parameter sets and evaluate amplifier performance with and without circuit parameter variance using transient circuit simulations. 
This approach inherently takes into account microwave reflections, unwanted mixing products, imperfect phase-matching, pump depletion, etc. 
In the case of RPM the resonance frequency spread is critical, while PCM is much less sensitive to parameter spread. 
We discuss degrees of freedom to make the JTWPA circuits more tolerant to parameter spread. 
Finally, our analysis shows that the flux-bias point where rf-SQUIDs exhibit Kerr-free nonlinearity 
is close to the sweet spot regarding critical current spread. 
\end{abstract}


\section{Introduction}
\IEEEPARstart{J}{osephson} 
traveling-wave parametric amplifiers (JTWPAs) hold great promise  for wideband amplification of few-photon-level signals at microwave frequencies. 
Their key characteristics are bandwidths of several GHz, added noise close to the quantum-limit, and power-handling capabilities above -100~dBm. 
These advantages are particularly important for simultaneous readout of many qubits \cite{Heinsoo2018,Krinner19}, or for reading out large arrays of kinetic inductance detectors\cite{Day2003}. 

JTWPAs utilize the nonlinear inductance of Josephson junctions, arranged as transmission line arrays of either single Josephson junctions \cite{White2015,Macklin2015,Feng2020,Qiu2022preprint}, dc superconducting quantum interference devices (SQUIDs) \cite{Planat2020}, rf-SQUIDs\cite{Zorin2016}, or superconducting nonlinear asymmetric inductive elements (SNAILs)\cite{Frattini2017,Zorin2017, Ranadive2022,Perelshtein22}. 
Parametric amplification occurs, depending on the order of nonlinearity, in the three-wave-mixing (3WM) or four-wave-mixing (4WM) regime, with $f_p=f_s+f_i$ and $2f_p=f_s+f_i$, respectively, where 
$f_p,f_s$, and $f_i$ are the pump,  signal, and  idler frequencies \cite{Agrawal}.  
If their respective wave numbers $k_p,k_s$, and $k_i$ fulfill the phase-matching criterion, $\Delta k=k_p-k_s-k_i=0$ for 3WM, 
high signal gain can be achieved, scaling exponentially with the circuit length \cite{Agrawal}. 


In the design of a 3WM JTWPA a major challenge is to ensure phase-matching along the whole length of the JTWPA.
This has been achieved in both 3WM and 4WM JTWPAs either by the resonant-phase-matching (RPM) technique \cite{OBrien2014} with lumped element resonators \cite{Macklin2015,  Feng2020, Qiu2022preprint} or distributed resonators \cite{White2015, Perelshtein22}, or by the periodic variation of the transmission line parameters \cite{Eom2012, Erickson2017, Planat2020, Malnou2021, Roudsari2022arx}, e.g., by periodic capacitance modulation (PCM). 
Both dispersion engineering methods create a narrow-band nonmonotonic feature in the dispersion relation $k(\omega)$, which is used to compensate the phase-mismatch $\Delta k$ by an appropriate choice of $f_p$. 
An additional challenge is to suppress unwanted parametric processes generating higher-frequency modes, which has been solved by increased dispersion of the transmission line for higher frequencies $f>f_p$ \cite{Malnou2021, Perelshtein22}.

Aside from these design challenges, any practical JTWPA implementation, made up of thousands of circuit elements, must be robust enough not to be affected by the expected parameter spread considering a typical fabrication process. 
Parameter variation impacts JTWPA performance in many ways, including phase-matching, impedance-matching, nonlinearity, etc. 
Previous works analyzed the effect of deteriorated phase-matching by resonance frequency spread in RPM-JTWPAs both analytically \cite{OBrien2014} and via Monte Carlo simulations using a linearized model\cite{Feng2020}, 
the tolerable critical current spread 
using analytical approaches\cite{Bell-Samolov2015,Zorin2019} and via quantum input-output theory \cite{Peng2022}, and microwave reflections due to parameter spread and point defects in a JTWPA using circuit simulations \cite{OPeatain2021preprint}. 
In this paper, we investigate both dispersion engineering concepts, RPM and PCM, 
and rigorously analyze their parameter spread sensitivity, employing transient circuit simulations in WRspice as described in \cite{Gaydamachenko2022}.
While most of the aforementioned works treated only one aspect of parameter spread isolatedly, our approach inherently includes the effect of spread on phase-matching, pump depletion, deviations from optimal bias point, microwave reflections, etc., and therefore gives a more realistic picture.

\section{Circuit Description and Analysis}

\begin{figure}[!t]
\centering
\includegraphics[width=3.5in]{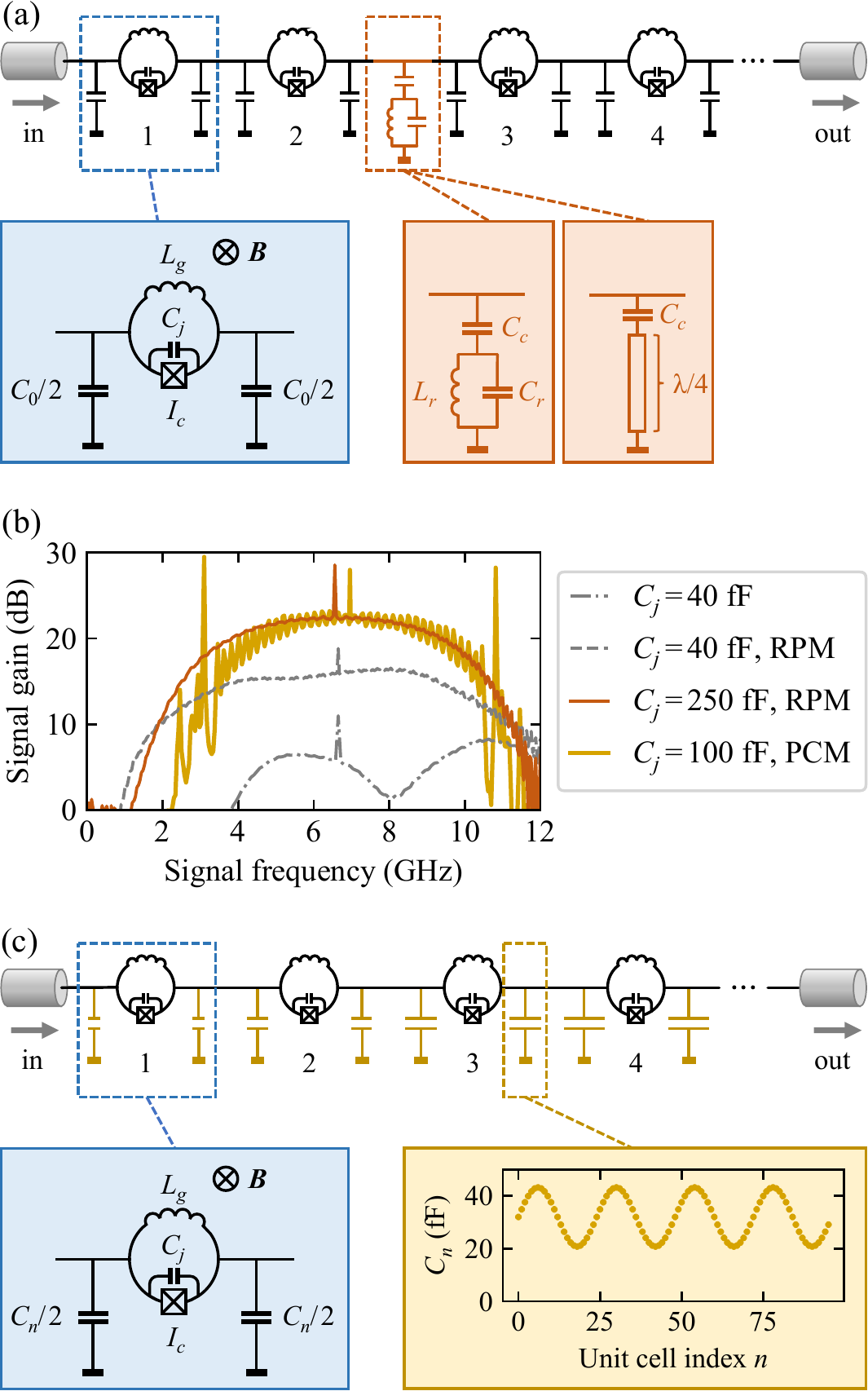}
\caption{JTWPA circuit model and effect of phase-matching on signal gain. 
(a) Circuit schematic of the RPM-JTWPA. It shows one section of $m\!=\!4$  identical unit cells and one phase-matching resonator attached in the center. 
The blue-shaded panel shows the unit cell, containing a flux-biased rf-SQUID and two capacitances $C_0/2$ to ground. 
RPM is implemented using either LC-resonators or $\lambda/4$-resonators, each capacitively coupled to the transmission line (brown-shaded panels). 
The resonators create a nonmonotonic feature in the dispersion relation $\omega(k)$ to achieve phase-matching. 
(b) Gain vs frequency profiles, obtained by transient circuit simulations for no dispersion engineering (dash-dotted), RPM (dashed line), RPM with increased effective SQUID capacitance $C_j$ (brown), and PCM using periodic capacitance modulation [gold, see panel (c)]. 
Phase-sensitive amplification at half of the pump frequency 
causes the distinctive peaks in the centers of all gain profiles. 
(c) Circuit schematic for periodic capacitance modulation (PCM). 
The values of the ground capacitances $C_n/2$, where $n$ is the unit cell index, are periodically varied. 
This opens a stopband in $\omega(k)$, which is used for phase-matching. 
Table~\ref{tab1} summarizes the circuit parameters. Simulations were performed with pump current of 6.59~µA for the RPM-JTPWA and 4.12~µA for the PCM-JTWPA,  see main text,  pump frequencies 13.3, 13.3, 13.12 and 13.92~GHz in the order of the legend, bias current of 7.5~µA, and signal current of 0.01~µA. 
}
\label{fig_1}
\end{figure}

We investigate a JTWPA architecture with rf-SQUIDs as nonlinear inductive elements in a transmission-line arrangement\cite{Zorin2016}. The rf-SQUIDs are nonhysteretic, i.e., having screening parameter $\beta_L=2\pi L_gI_c/\Phi_0<1$, where $L_g$ is the geometrical inductance, $I_c$ is the critical current of the Josephson junction, and $\Phi_0$ is the flux quantum. 
Each rf-SQUID forms a unit cell with two capacitors $C_0/2$ to ground, see Fig.~\ref{fig_1}, yielding impedance $Z_0=\sqrt{L_g/C_0}=50~\Omega$. 
A magnetic field $\boldsymbol{B}$, or alternatively a dc current $I_{dc}$, 
flux-biases the rf-SQUIDs such that the noncentrosymmetric $\chi^{(2)}$-type nonlinearity of the circuit is large and the $\chi^{(3)}$-type Kerr nonlinearity vanishes \cite{Zorin2016}. 
At this flux bias, the effective linear rf-SQUID inductance is $L_S=L_g$. 
The $\chi^{(2)}$-type nonlinearity facilitates parametric amplification in the 3WM 
regime when driven by a strong pump wave. 
Taking into account the Josephson junction capacitance $C_j$, the resulting dispersion relation \cite{Zorin2019}, 
\begin{equation}
    k(\omega) =  \frac{2}{a} \arcsin \left( \frac{\omega \sqrt{L_gC_0}}{2\sqrt{1-\omega^2 L_gC_j}} \right),
    \label{k-vs-w}
\end{equation}
where $a$ is the physical length of a unit cell, causes a nonzero phase mismatch $\Delta k>0$. 
A possible way to compensate $\Delta k$ is lowering $k_p$ with the help of either RPM or PCM. 
The effect of adding RPM elements to the JTWPA circuit is illustrated by the dashed and the dash-dotted lines in Fig.~\ref{fig_1}b, corresponding to the JTWPA with and without RPM and otherwise identical parameters. 
Further improvement aims at suppressing unwanted parametric processes. 
To achieve this, the effective SQUID capacitance $C_j$ is enlarged, e.g., by adding a capacitor in parallel to the Josephson junction, lowering the SQUID plasma frequency and increasing the dispersion, eq.~(\ref{k-vs-w}), which causes phase mismatch for the higher-frequency mixing processes \cite{Zorin2021}. 
This also increases $\Delta k$, which can be compensated by RPM, however. 
The effectiveness of this approach is demonstrated by the solid brown line in Fig.~\ref{fig_1}b, indicating almost pure 3WM and yielding $>\,\,$20~dB gain in a 3-dB-bandwidth of more than 5~GHz.  
Increasing $C_j$ above 250~fF does not result in significantly higher gain but causes stronger gain ripple due to the alteration of the line impedance \cite[eq.~(C8)]{Dixon2020}. 
The circuit parameters used in the simulations are given in Table~\ref{tab1}.

\begin{table}[b]
\begin{center}
\caption{Circuit parameters used for transient circuit simulations}
\label{tab1}
\begin{tabular}{ c  c  c }
\toprule
\textbf{}& \textbf{JTWPA with RPM} & \textbf{JTWPA with PCM}\\
\midrule
Inductance & $L_{g}=80~\text{pH}$& $L_{g}=80~\text{pH}$ \\
Critical current & $I_{c}=1.03~\text{µA}$& $I_{c}=1.03~\text{µA}$ \\
Junction capacitance & $C_{j}=250~\text{fF}$& $C_{j}=100~\text{fF}$ \\
Screening parameter & $\beta_{L}=0.25$ & $\beta_{L}=0.25$ \\
\midrule
Resonator capacitance & $C_{r}= 1~\text{pF}$ & \\
Resonator inductance & $L_{r}= 150~\text{pH}$ & \\
Coupling capacitance & $C_{c}= 20~\text{fF}$ & \\
Unit cells per resonator & $m= 4$ & \\
\midrule
Ground capacitance & $C_0=32~\text{fF}$ & $C_0=32~\text{fF}$ \\
Modulation depth &  & $\zeta=0.3$\\
Unit cells per period & & $m= 24$ \\
\midrule
Number of unit cells & $N= 1000$ & $N=1488$\\
\toprule
\end{tabular}
\end{center}
\end{table}

To implement PCM, we periodically vary the ground capacitances $C_n$ 
with a modulation depth $\zeta$ and a period $m$, 
\begin{equation}
    C_n=C_0\left(1+\zeta\sin(2\pi n/m)\right),
\end{equation}
where $C_0$ is the mean value, see Fig.~\ref{fig_1}c. 
The PCM opens a stopband in $k(\omega)$ at $\omega_1 = \pi/(m\sqrt{L_gC_0})$ and fulfills the same purpose as the resonant feature in RPM. 
The combination of PCM and the SQUID plasma resonance leads to effective suppression of unwanted modes \cite{Perelshtein22} and simultaneously provides phase-matching for 3WM. 
In contrast to JTWPAs based on single Josephson junctions, where $I_p<I_c$, no such limit exists for rf-SQUID based JTPWAs, where the Josephson junction is shunted by a superconducting inductance. 
Instead, the phase drop across a Josephson junction should be kept below $\sim$1~rad \cite{Zorin2016,Zorin2021}, above which the circuit may behave unpredictable. 
Therefore, we chose a pump amplitude of 0.8~rad for the RPM-JTWPA, which translates to a pump current $I_p=\phi_p\Phi_0/2\pi L_g=6.59$~µA. 
Due to the superposed forth- and back-propagating components of the pump wave close to the stopband \cite{Planat2020}, 
for the PCM-JTWPA we chose 0.5~rad for the pump amplitude, translating to $I_p=4.12$~µA, and increase the number of unit cells to achieve the same gain. 
The solid ocher line in Fig.~\ref{fig_1}b shows the gain profile of the PCM-JTWPA,   
where the increased ripple stems from sidelobes of the stopband and is an unwanted side effect of PCM \cite{Gaydamachenko2022}. 
A higher-order stopband 
created by PCM causes the two distinctive peaks at both edges of the gain profile, confining the usable range of signal frequencies. 
The peaks in the center of each gain profile indicate 
phase-sensitive amplification \cite{Perelshtein22,Gaydamachenko2022}, an interesting regime at $f_s=f_i=f_p/2$ which is experimentally accessible in 3WM-JTWPAs.

\section{Parameter Spread Sensitivity}

\begin{figure*}[!t]
\centering
\includegraphics[width=7.16in]{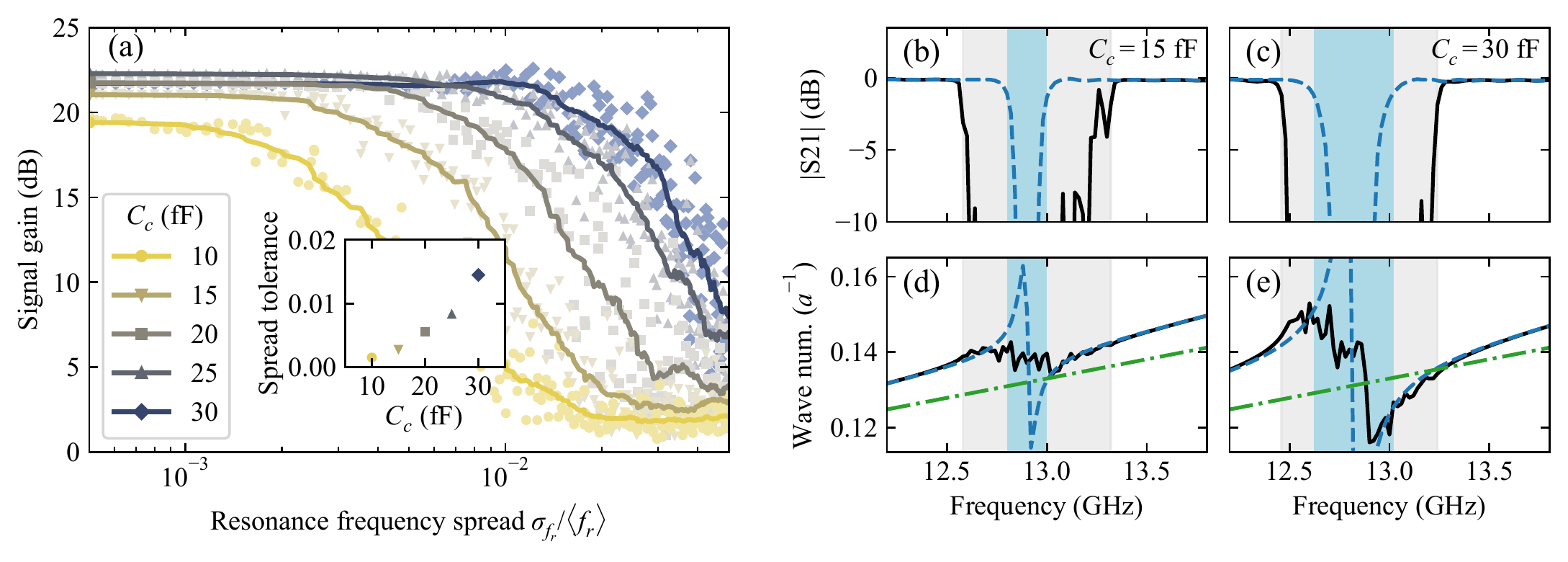}
\caption{Sensitivity of the RPM-JTWPA to resonance frequency spread. (a) Power gain as a function of normalized resonance frequency standard deviation for five coupling capacitances. 
The resonance frequencies of all 250 individual resonators 
are drawn from a normal distribution with standard deviation $\sigma_{f_r}$ and  mean value $\langle f_r\rangle$. 
For each $\sigma$ the maximum gain within a range of pump frequencies is plotted, see main text. 
As $\sigma_{f_r}$ increases, the gain drops due to impaired phase-matching along the transmission line.  
Solid lines are moving-averages of 20 samples for illustrative purpose. 
The inset shows the value of $\sigma_{f_r}/\langle f_r\rangle$, where the moving average gain curve decreases by 1 dB. Larger coupling capacitance $C_c$ results in greater robustness against resonator frequency spread. 
(b)--(e) Small-signal transmission of the RPM-JTWPA in the vicinity of the resonance frequency with and without resonator spread. 
Dashed blue lines depict the zero-spread cases, and the blue-shaded regions indicate the corresponding resonance dips, where $|S_{21}|\!<\!-1$~dB. 
The dash-dotted green lines in (d) and (e) show the dispersion-free wave-number. 
Phase-matching is achieved when the dashed and the dash-dotted lines intersect outside the resonance dip, avoiding severe pump depletion. This is possible for both shown values of $C_{c}$ at zero spread. 
The solid black lines are typical examples with $\sigma_{f_r}/\langle f_r\rangle$=1.1\%, illustrating the cases with resonance frequency spread. 
The resonance frequency spread results in a widened resonance dip, shown by the grey-shaded area, and also in weakened modification of the wave number. 
(d) In the case of the lower coupling strength there is no intersection of the solid line with the dash-dotted line, resulting in reduced gain, see (a).  
(e) For stronger coupling, phase-matching is still possible and the frequency of zero phase-mismatch lies outside  the resonance dip. 
All simulations of (a) were performed with an identical pump current of $I_p=6.59$~µA for comparability, and with a signal frequency of 7.7~GHz and pump frequencies in the range of 12.98--14.0~GHz. The small-signal simulations of (b)--(e) were carried out with a current of 0.1~µA~$<<I_p$. 
}
\label{fig_2}
\end{figure*}

Next, we modify the JTWPA circuit models to incorporate unit cell to unit cell variation to analyze the effect of variation of the parameters $I_c, L_g, C_{n,j,c}$, and $f_r$ on the performance of the JTWPA. 
In the case of the resonance frequency spread, 
we vary all capacitances $C_r$ in the RPM-JTWPA circuit, having a normal distribution with a mean value $\langle C_r \rangle$ equal to the nominal value, and with a standard deviation $\sigma_{C_r}$ being increased in small steps. 
Keeping inductance $L_r$ constant, $\sigma_{L_r}=0$, the resonance frequency spread is \cite{Goodman1960}
\begin{equation}
    \frac{\sigma_{f_r}}{\langle f_r \rangle} = \frac{1}{2} \sqrt{
    \frac{\sigma_{L_r}^2 \sigma_{C_r}^2}{\langle L_r \rangle^2 \langle C_r \rangle^2} + 
    \frac{\sigma_{L_r}^2}{\langle L_r \rangle^2} + \frac{\sigma_{C_r}^2}{\langle C_r \rangle^2}}
    =\frac{1}{2} \frac{\sigma_{C_r}}{\langle C_r \rangle}.
    \label{eq_fr_spread}
\end{equation}
For each value of $\sigma_{C_r}$, in the following referred to as sample, a set of simulations is carried out for a range of pump frequencies $f_p$. 
This is necessary due to the stochastic nature of phase-matching in a circuit with parameter spread; the frequency where the phases are matched differs from sample to sample, c.f. Fig.~\ref{fig_2}e. 
Each data point in Fig.~\ref{fig_2}a is the highest achieved gain of one sample, and all samples in the ensemble are statistically independent.  
To visualize the trend of the achievable gain depending on $\sigma_{C_r}$, we calculate a moving average with a window size of 20 samples.

To define spread tolerance, we take the values of $\sigma_{f_r}/\langle f_r \rangle$ for which the moving average gain drops by 1~dB from its zero-spread value. 
The inset in Fig.~\ref{fig_2}a shows the spread tolerance for five different coupling capacitances and indicates generally increased robustness for larger $C_c$.   
To illustrate this, Fig.~\ref{fig_2}b--e present the small-signal transmission in the vicinity of the resonance dip of samples having 2\% capacitance spread and 1\% inductance spread, yielding $\sigma_{f_r}/\langle f_r \rangle=1.1\text{\%}$ using (\ref{eq_fr_spread}), and reflecting realistic values for lumped element resonators \cite{Macklinthesis,Tolpygo2015_inductance,Tolpygo2021,Cicak2010}. 
Resonance frequency spread results in a widened resonance dip and in reduced modification of the wave number. Whether both phase-matching and avoiding severe pump depletion are simultaneously possible depends  on the coupling strength. 
Apart from stronger coupling, the resonator spread tolerance can be increased by increasing the number of resonators in the circuit, i.e., to decrease period $m$. For $m=1$, resonator spread tolerances of $>2\text{\%}$ were reported \cite{OBrien2014,Feng2020}. This comes at the cost of larger footprint on chip and higher complexity. 
Furthermore, 
instead of lumped element resonators, distributed resonators can be used, exhibiting much lower spread $<0.1$\%\cite{Mazinthesis}, also at the cost of larger footprint, c.f. \cite{White2015}. 
Note that 3WM, when compared to 4WM, requires only half of the length of RPM resonators due to the twice higher pump frequency.

\begin{figure*}[!t]
\centering
\includegraphics[width=7.16in]{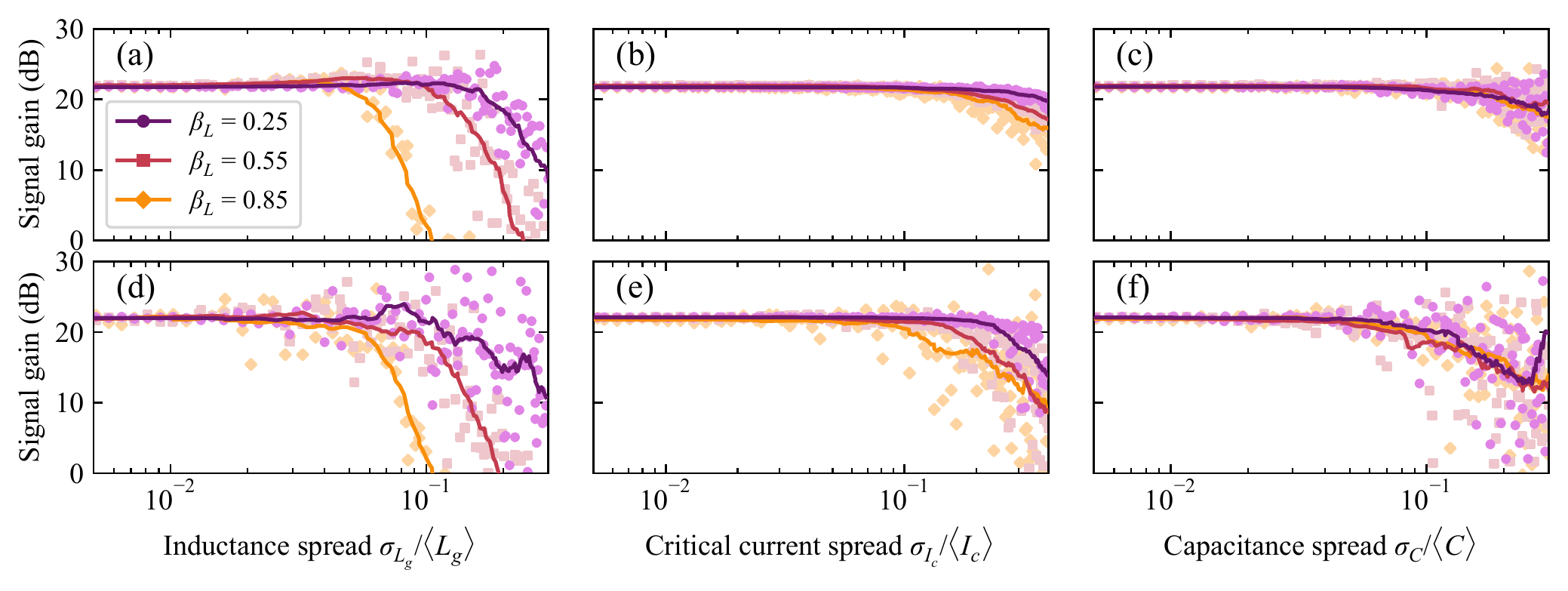}
\caption{
Spread sensitivity of the RPM-JTWPA (a)--(c) and the PCM-JTWPA (d)--(f) for three screening parameter values $\beta_{L}$. 
In each panel, the  respective parameter is varied according to a normal distribution, while the remaining parameters are kept constant. Capacitance spread refers to the simultaneous variation of the capacitances $C_n$, $C_j$, and $C_c$. 
Solid lines are moving-averages of 20 adjacent samples to illustrate the trends.  
Increasing parameter spread leads to a drop of the gain, and to microwave reflections, visible as strongly diverging gain values higher than the zero-spread gain. 
Even though the PCM-JTWPA is  more sensitive to parameter spread than the RPM-JTWPA, it does not contain a critical element like the resonators in RPM. 
Both RPM and PCM are more sensitive to inductance spread than to critical current spread or capacitance spread, and lower values of  $\beta_{L}$ are favourable. 
The simulations of (a)-(c) were performed with a signal frequency of 7~GHz, and with pump frequencies in the range of 12.98--13.1~GHz and pump currents of 6.59, 2.8, 1.8~µA for $\beta_L\!=$ 0.25, 0.55, 0.85, respectively. The simulations of (d)-(f) were performed with a signal frequency of 7.7~GHz, and with pump frequencies in the range of 12.98--13.1~GHz and pump currents of 3.84, 1.63, 1.07~µA for $\beta_L\!=$ 0.25, 0.55, 0.85, respectively.
}
\label{fig_3}
\end{figure*}

Although the RPM-JTWPA is vulnerable to resonance frequency spread, the spreads of the other parameters, 
$L_g,I_c$, and $C_{0,c,j}$, 
play a minor role, see Fig.~\ref{fig_3}a--c, because they do not directly impact phase-matching. 
This is different in the case of PCM, see Fig.~\ref{fig_3}d--f. 
The stopband, providing the means for phase-matching in PCM, is facilitated by the spatially periodic form of quantities like the wave impedance or the wave velocity. 
Consequently, stochastic spatial variations to that periodic shape impact the characteristics of the stopband and, hence, phase-matching. 
On the other hand, within one period, consisting of many unit cells, deviations cancel each other out, which is in great contrast to RPM, where each individual resonator directly impacts phase-matching. 
Leaving the spread of the resonance frequencies aside, PCM is 2--4 times more sensitive to the spread of $L_g,I_c,C_{n,j}$ than RPM. 
However, the spread sensitivity analysis shows that typical spread values, being on the order of $<2$\% for capacitances and inductances\cite{Tolpygo2015_inductance,Tolpygo2021,Cicak2010}, and 1--8\% for the critical currents of both Nb and Al Josephson junctions \cite{Tolpygo2015_JJ,Bumble2009,Pop2012,Kreikebaum2020,Osman2021}, are tolerable. 
Taking this into account, we conclude that PCM does not have a critical circuit element as the resonators in RPM.

Among the parameters $L_g,I_c,C_{0,c,j}$, the impact of inductance spread is the strongest, since $L_g$ directly influences the flux biasing, inductance, and nonlinearity of the rf-SQUIDs and, consequently, the local impedance and  wave number of the line. 
This does not only deteriorate phase-matching and reduce the signal gain, but also causes increased microwave reflection\cite{OPeatain2021preprint} and gain ripple. 
Microwave reflections at inhomogeneities in the circuit interfere either destructively or constructively, leading to a dip below or a peak beyond the zero-spread gain. 
Such diverging gain values are most pronounced  in Fig.~\ref{fig_3}d, but can be also seen in the other panels for large spread values. 
Moreover, Fig.~\ref{fig_3}a,d show that the inductance spread tolerance strongly depends on parameter $\beta_L\propto 2L_gI_c$, which is a degree of freedom in rf-SQUIDs. 
To vary $\beta_L$, we keep $L_g$ constant and increase $I_c$. Since the exponential gain coefficient of the JTPWA is proportional to $\beta_L$\cite{Zorin2016}, we reduce the pump amplitude in the simulation to achieve similar  gain for all $\beta_L$.

Interestingly, the rf-SQUID based JTWPA is quite insensitive to critical current spread 
up to spreads of 10--20\%, see Fig.~\ref{fig_3}b,e, 
in contrast to 3\% for the SNAIL-based reversed-Kerr JTWPA\cite{Bell-Samolov2015}. 
This is because the flux bias where the Kerr-like nonlinearity of the rf-SQUID vanishes\cite{Zorin2016} is close to the sweet spot of critical current spread tolerance. 
At that point, the linear inductance of the Josephson junction goes to infinity, so the effective linear rf-SQUID inductance is $L_S=L_g$. Hence, as a first-order approximation, the Josephson junctions contribute only to the nonlinear but not to the linear characteristics of the circuit. 
Using $\beta_L=0.25$ and taking into account an inductance spread of 1\% and a critical current spread of 5\%, the resulting SQUID inductance spread, see (\ref{eq_Ls_spread}) in appendix,  is 1.4\%, and is determined almost entirely (98\%) by inductance spread.

\section{Conclusion}

In summary, we analyzed the sensitivity of JTWPA gain on the spread of critical currents, inductances, capacitances, and resonance frequencies, using transient circuit simulations. 
In the case of a JTWPA with RPM, the resonance frequency spread is critical to the performance of the amplifier; how much resonance frequency spread can be tolerated is determined by the resonator coupling strength. 
Apart from the resonator spread, the RPM-JTWPA is more tolerant than the PCM-JTWPA to the spread of the other circuit parameters. 
Among these, the inductance spread has the largest impact on JTWPA performance, depending on the SQUID screening parameter. 
The Kerr-free rf-SQUID based JTWPA is rather insensitive to Josephson junction critical current spread. 
Our analysis can be employed also to other JTWPA architectures, and the results should also be applicable to SNAIL-based JTWPAs in particular. 
We think that understanding the sensitivity of a JTWPA concept to parameter spread is key to its optimization and practical realization, and we believe that our results help paving the way to a practical implementation of rf-SQUID based JTWPAs.

\appendix
The linear inductance of an rf-SQUID is given by
\begin{equation}
    L_S=\frac{L_g}{1+\beta_L\cos{\phi_{dc}}}
\end{equation}
and the dc phase $\phi_{dc}$ is the root of equation \cite{Likharevbook} 
\begin{equation}
    \phi_{dc}+\beta_L\sin{\phi_{dc}}-\phi_e=0,
    \label{eq-transcendental}
\end{equation}
with the phase $\phi_e=2\pi L_gI_{dc}/\Phi_0$, given by a dc bias current $I_{dc}$. The latter is chosen such that $\phi_e=\beta_L+\pi/2$ and $\phi_{dc}=\pi/2$, yielding a zero Kerr nonlinearity coefficient\cite{Zorin2016}.

To show why the rf-SQUID inductance $L_S$ is relatively insensitive to critical current spread, we propagate the variances $\sigma_{L_g}^2$ and $\sigma_{I_c}^2$ to the 
variance of $L_S$, 
\begin{equation}
    \sigma_{L_S}^2 \approx 
    \left| \frac{\partial L_S}{\partial L_g} \right|^2 \sigma_{L_g}^2 + \left| \frac{\partial L_S}{\partial I_c} \right|^2 \sigma_{I_c}^2\;. 
    \label{eq_Ls_propagation}
\end{equation}
Bearing in mind $\beta_L=2\pi L_gI_c/\Phi_0$, and taking the implicit derivatives of the transcendental equation (\ref{eq-transcendental}), 
\begin{equation}
    \frac{\partial \phi_{dc}}{\partial L_g}=\frac{1}{L_g} \frac{\phi_e-\beta_L\sin\phi_{dc}}{1+\beta_L\cos\phi_{dc}},
\end{equation}
\begin{equation}
    \frac{\partial \phi_{dc}}{\partial I_c}=-\frac{1}{I_c} \frac{\beta_L\sin\phi_{dc}}{1+\beta_L\cos\phi_{dc}},
\end{equation}
we get the derivatives
\begin{multline}
    \frac{\partial L_S}{\partial L_g}=\\
    \frac{(1+\beta_L\cos\phi_{dc})-\beta_L\left(\cos\phi_{dc}-\sin\phi_{dc} \frac{\phi_e-\beta_L\sin\phi_{dc}}{1+\beta_L\cos\phi_{dc}}\right)} {(1+\beta_L\cos\phi_{dc})^2}
    \label{eq-dLsdLg}
\end{multline}
and
\begin{equation}
    \frac{\partial L_S}{\partial I_c} = -\frac{L_g\beta_L}{I_c}\frac{\left(\cos\phi_{dc} + \sin\phi_{dc}\frac{\beta_L\sin\phi_{dc}}{1+\beta_L\cos\phi_{dc}}\right)} {(1+\beta_L\cos\phi_{dc})^2}.
    \label{eq-dLsdIc}
\end{equation}
Evaluating (\ref{eq-dLsdLg}) and (\ref{eq-dLsdIc}) at $\phi_{dc}=\pi/2$ and using (\ref{eq_Ls_propagation}) yields the normalized standard deviation of the rf-SQUID inductance
\begin{equation}
    \frac{\sigma_{L_S}}{\langle L_S \rangle} = \sqrt{
    \left(1+\frac{\pi}{2}\beta_L\right)^2\frac{\sigma_{L_g}^2}{\langle L_g \rangle^2} + \beta_L^4
    \frac{\sigma_{I_c}^2}{\langle I_c \rangle^2}}\;. 
    \label{eq_Ls_spread}
\end{equation}

\section*{Acknowledgment}
The authors would like to thank D.~Hanisch and F.~Feldhoff for fruitful discussions, and  T.~Dixon, D.~Müller and V.~Rogalya for their help in setting up the simulator platform.

\bibliography{JTWPA_references}{}
\bibliographystyle{IEEEtran}

\vfill

\end{document}